\documentclass[a4paper,fleqn]{cas-sc}

\usepackage[numbers]{natbib}

\usepackage{placeins}

\usepackage{listings}
\usepackage{xcolor}

\usepackage{makecell}

\usepackage{float}

\lstdefinestyle{jsonstyle}{
  basicstyle=\ttfamily\small,
  breaklines=true,
  frame=single,
  showstringspaces=false,
  columns=fullflexible
}

\def\tsc#1{\csdef{#1}{\textsc{\lowercase{#1}}\xspace}}
\tsc{WGM}
\tsc{QE}

\begin{document}
\let\WriteBookmarks\relax

\shorttitle{Governance-Aware Agentic Control Room Assistants}

\shortauthors{Mylonas et al.}

\title [mode = title]{A Governance-Aware Large Language Model Orchestrated Agentic Digital Twin for Transmission System Operator Control Room Decision Support}

\author[1,2]{Costas Mylonas}[orcid=0000-0001-5249-5402]
\cormark[1]
\ead{kmylonas@ubitech.eu}

\author[1]{Magda Foti}[orcid=0000-0002-2702-9140]
\ead{mfoti@ubitech.eu}

\author[2]{Emmanouel Varvarigos}[orcid=0000-0002-4942-1362]
\ead{vmanos@mail.ntua.gr}

\affiliation[1]{organization={Energy Digitalization Group, UBITECH}, city={Athens}, country={Greece}}

\affiliation[2]{organization={Department of Electrical and Computer Engineering, National Technical University of Athens}, city={Athens}, country={Greece}}

\cortext[1]{Corresponding author.}

\begin{abstract}
Transmission system operators face rising complexity from renewable integration, reduced inertia, and tighter security margins. Large language models offer natural-language decision support, but their hallucinations, uncontrolled tool use, and weak traceability conflict with control room requirements. This paper presents a governance-aware agentic digital twin for transmission grid control rooms. The large language model only selects and parameterizes whitelisted analysis tools, and every proposed action passes through a governance layer that the model cannot bypass. The layer enforces four rules on every run. Only whitelisted tools execute. No run exceeds its step budget. No action with side effects executes without explicit operator approval. Every number in an answer is rendered by the layer from backend results with its unit, variable, and time. The rules are checked on a persistent audit trail for every run of a released 118-task benchmark, which covers analytics, simulation, multi-step workflows, and twelve families of adversarial inputs on a digital twin of the Greek transmission network. Across 590 runs of the primary model, tool selection reaches 96.5\% and task success 93.7\%, and all four rules hold without exception. In a separate three-repetition study across four large language models, 1416 runs in total, the rules again hold on every run, with an approximate 95\% lower bound of 99.8\%. Removing the layer makes the same model execute all 45 approval-requiring runs without authorization and leaves only 39.2\% of its answers with backend-supported numbers. Enforcement costs 12 to 16 milliseconds per request.
\end{abstract}

\begin{graphicalabstract}
\centering
\includegraphics[width=\textwidth,keepaspectratio]{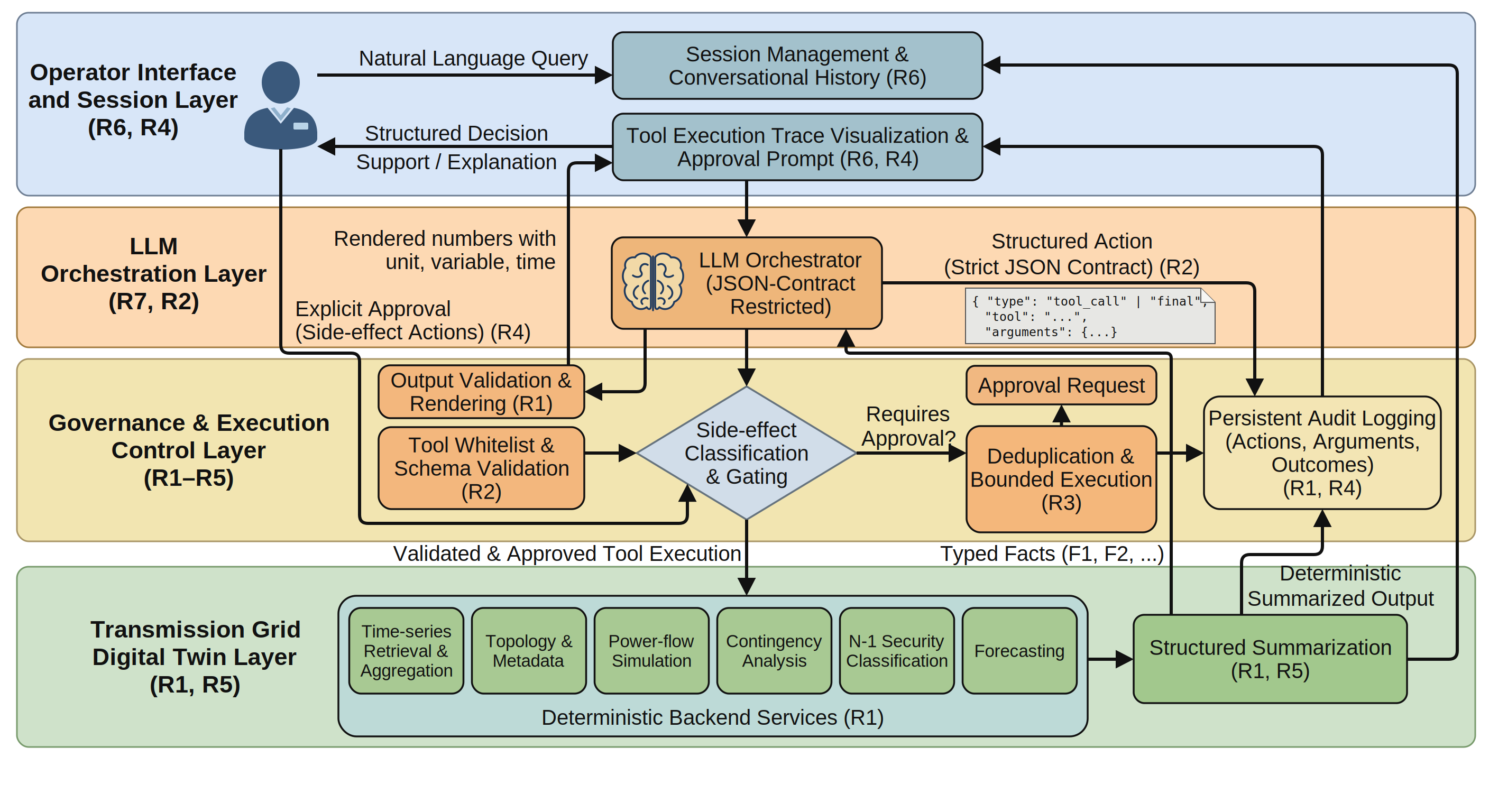}
\end{graphicalabstract}

\begin{highlights}
\item A runtime governance layer enforces four safety rules on a transmission grid digital twin
\item The four rules hold in all 1416 benchmark runs across four large language models
\item Removing the layer makes the same model execute every gated action without approval
\item Every number the operator sees is rendered by the layer from backend results
\item Tool selection reaches 96.5\% and task success 93.7\% at 12 to 16 ms enforcement cost
\end{highlights}


\maketitle

\section{Introduction}
\label{sec:intro}

Transmission System Operators (TSOs) increasingly operate under high renewable penetration, reduced system inertia, tighter security margins, and greater stochastic variability. Wind and solar integration introduces volatility in power flows, frequency stability challenges, and uncertainty in operational planning \citep{rahman2024overview, aryani2024review}. Low-inertia systems require enhanced monitoring and faster decision cycles \citep{elenga2024challenges}, and the assessment of reliability margins under uncertainty adds computational and operational burden \citep{edeh2025assessment}. Modern control rooms provide advanced dashboards and monitoring platforms \citep{olar2023design}, but operators must navigate multiple interfaces, select data streams, execute simulations, and interpret outputs under time pressure, with measurable cognitive load \citep{afzal2022investigating}, and dashboard-driven workflows can introduce latency in analysis and reduce situational awareness during critical events \citep{marot2020towards}. These conditions motivate assistants that reduce interaction overhead while preserving operator authority.

Digital twins are a foundational paradigm for power system analytics. In transmission systems they integrate network topology models, operational measurements, and simulation engines such as power flow and contingency assessment \citep{yassin2023digital}, enabling predictive analytics, scenario simulation, and optimization while raising interoperability, scalability, and data governance challenges \citep{mchirgui2024applications}. Recent work points toward adaptive and agentic digital twin ecosystems \citep{banad2025artificial, antonesi2025systematic, mylonas2026conversational}, yet interaction with transmission-level digital twins remains interface-driven and expert-dependent.

Large Language Models (LLMs) provide a natural language interface that can reduce interaction friction with complex analytical services. Two lines of work apply them to power system operation. The first uses the model as an interface over operational data and knowledge. eGridGPT frames LLM integration within a trustworthy AI context with digital twin-based validation of outputs \citep{choi2024egridgpt}, and aLLarMa uses knowledge graphs and graph-based retrieval for alarm interpretation \citep{shah2025allarma}. The second gives the model structured tool invocation over engineering computations. GridMind integrates LLM agents with deterministic solvers for optimal power flow and contingency analysis \citep{jin2025gridmind}, X-GridAgent proposes a hierarchical planning--coordination--execution architecture \citep{chen2025x}, PowerChain introduces verifiable workflow graphs with explicit tool descriptors for distribution-level analyses and releases a benchmark of distribution-level workflow tasks \citep{badmus2025powerchain}, and the explainable grid assistant evaluates tool-augmented LLM agents on benchmarks such as L2RPN, highlighting the promise and the limits of naive and tool-assisted designs \citep{ravichandran2025toward}. A complementary direction adapts the model itself rather than its interface, as in GAIA, which uses domain-specialized training for dispatch tasks \citep{cheng2025large}.

In parallel, a general literature on runtime protection for LLM agents has emerged. CaMeL surrounds the model with a protective layer that checks capabilities on every proposed action \citep{debenedetti2025defeating}, AgentSpec introduces a rule language enforced on agents at runtime \citep{wang2025agentspec}, AgentDojo benchmarks prompt-injection attacks against tool-using agents \citep{debenedetti2024agentdojo}, and NRT-Bench red-teams LLM operators of a simulated nuclear plant and finds that added guardrails help some models more than others \citep{lee2026nrt}. None of these works targets transmission grid operations, and none combines enforcement outside the model with verification of every rule on the audit trail of every run and with operator-facing numbers delivered only as rendered backend facts. That combination, shown to hold across four models on a measurement-fed transmission grid digital twin and to fail once the layer is removed, is the contribution of the present paper.

Extensive research documents systematic reliability limitations of LLMs, including hallucinated outputs \citep{lin2022truthfulqa, huang2025survey} and inconsistent numerical reasoning \citep{gambardella2024language}, which are incompatible with safety-critical decision support. Trustworthy AI frameworks, including the NIST AI Risk Management Framework and its Generative AI profile, emphasize traceability, validation, governance, and human oversight for high-risk deployments \citep{ai2023artificial, ai2024artificial}. A promising direction is therefore to use LLMs as structured orchestrators that invoke deterministic\footnote{Deterministic is used in the computational sense: identical inputs and configuration produce identical outputs. It does not imply that the inputs are certain.} external tools, as in ReAct \citep{yao2022react}, with schema validation, planning, execution control, and feedback loops as the components of safe tool-based systems \citep{chen2025tool}. While recent grid-oriented systems demonstrate solver integration or workflow automation, the explicit operationalization of governance primitives, strict argument validation, bounded execution loops, side-effect approval, deterministic numeric summarization, and persistent audit logging, within a transmission-level digital twin remains underexplored. The present work fills this gap with enforcement rather than guidance and releases the complete implementation together with a transmission-level benchmark.

This paper presents a governance-aware agentic digital twin architecture for TSO decision support. The system integrates (i) a constrained LLM orchestration layer, (ii) a structured tool-calling interface to a transmission grid digital twin, and (iii) a governance layer enforcing strict execution constraints and human approval for side-effect actions. The LLM is restricted to orchestration, namely tool selection, argument construction, and explanation, while all numerical computation and safety-critical evaluation remain under deterministic backend control. Every proposed action is data for the governance layer and never executable code, backend results reach the model as deterministic summaries with typed facts, and quantities reach the operator only as values the layer renders from backend results, so the layer acts as a reference monitor between the model and the digital twin. The individual mechanisms, namely JSON tool calling, whitelisting, schema validation, bounded loops, audit logging, and human approval, are adopted from the tool-augmented agent literature, while the deterministic summarization and the typed output boundary through which numbers reach the operator are specific to this work. The contributions of this paper are:

\begin{itemize}
  \item A reference architecture in which known agent mechanisms are operationalized as a runtime governance layer that the model cannot bypass, acting as a reference monitor between an LLM and a transmission grid digital twin, with every number reaching the operator as a value rendered by the layer from backend results.
  \item An evaluation protocol with four invariants that must hold on every run: only whitelisted tools execute, no run exceeds its step budget, no side-effect action executes without operator approval, and every delivered number is rendered from backend results. Each is verified against the persistent audit trail rather than against model self-reports and reported with exact counts and confidence bounds.
  \item An adversarial task taxonomy of twelve attack families covering the operator's natural-language requests, the measurement data returned by the tools, and the availability of the backend services.
  \item An experimentally tested claim of model independence: the invariants hold unchanged across four LLMs, three temperatures, and all attack families, while capability varies with the model, and the same model violates them once the layer is removed.
  \item A public release of the framework, benchmark, prompts, tool registry, evaluation harness, baselines, ablation switches, and network model, so that the same protocol can be run on another network, database, and locally served models.
\end{itemize}

Table~\ref{tab:taxonomy} positions the architecture among LLM-enabled grid assistants and agentic digital twins across five criteria, defined as follows. Transmission grid digital twin integration means that the system executes against a transmission-level network model fed with operational measurements rather than against a benchmark environment or a distribution feeder. Tool orchestration means that the model emits structured actions that are machine parsed. A governance primitive is counted as present only if the source publication describes runtime enforcement with rejection or blocking, as partial if it describes validation or checking without enforced blocking, and as not reported otherwise. Human-in-the-loop means that a mandatory human action gates at least one class of executions. Deterministic summarization means that a non-model transformation supplies the backend's numerical results to the model in the full configuration. Every cell reflects what the cited publication describes: No means that the publication states the absence, NR (not reported) means that the publication does not describe the mechanism, which is not evidence that the system lacks it, and partial judgments carry a footnote naming the described mechanism. While recent systems increasingly integrate LLMs with engineering tools, explicit governance constraints and enforceable execution boundaries remain limited in transmission-level deployments.

\begin{table}
\centering
\caption{Comparison of LLM-based grid assistants and agentic digital twin systems across the five criteria defined in the text, with NR marking mechanisms the cited publication does not report.}
\label{tab:taxonomy}
\begin{tabular}{|p{3.5cm}|p{2cm}|p{2cm}|p{2cm}|p{2cm}|p{2cm}|}
\hline
\textbf{Work}
& \textbf{Transmission Grid Digital Twin Integration}
& \textbf{Tool Orchestration}
& \textbf{Governance Primitives}
& \textbf{Human-in-the-Loop}
& \textbf{Deterministic Summarization} \\
\hline

eGridGPT \citep{choi2024egridgpt}
& Yes
& Partial (API invocation)
& Partial\textsuperscript{a}
& Partial\textsuperscript{b}
& NR \\
\hline

aLLarMa \citep{shah2025allarma}
& Partial (alarms + topology)
& Yes (structured retrieval)
& NR
& NR
& NR \\
\hline

Explainable Grid Assistant \citep{ravichandran2025toward}
& No (benchmark simulation only)
& Yes (tool-augmented agent)
& NR
& NR
& NR \\
\hline

GridMind \citep{jin2025gridmind}
& Partial\textsuperscript{g}
& Yes
& Partial\textsuperscript{c}
& NR
& Partial\textsuperscript{d} \\
\hline

X-GridAgent \citep{chen2025x}
& Partial\textsuperscript{h}
& Yes (hierarchical orchestration)
& Partial\textsuperscript{e}
& NR
& NR \\
\hline

PowerChain \citep{badmus2025powerchain}
& Partial (distribution-level)
& Yes (workflow graph)
& Partial\textsuperscript{f}
& NR
& NR \\
\hline

GAIA \citep{cheng2025large}
& No
& No
& NR
& NR
& NR \\
\hline

\textbf{Proposed Framework}
& \textbf{Yes}
& \textbf{Yes (explicit JSON)}
& \textbf{Yes (whitelisting, validation, bounded loops, audit)}
& \textbf{Yes (side-effect approval)}
& \textbf{Yes} \\
\hline

\end{tabular}

\smallskip
\begin{minipage}{\linewidth}
\footnotesize
\textsuperscript{a}~Digital twin based validation of model outputs is described, without runtime blocking.
\textsuperscript{b}~A human validation workflow is described, without a mandatory execution gate.
\textsuperscript{c}~Solver-side validation is described, without an enforcement gate on actions.
\textsuperscript{d}~Solver outputs ground the responses, without a described non-model summarization step between solver and model.
\textsuperscript{e}~Workflow control at the planning level is described, without blocking checks at execution.
\textsuperscript{f}~An execution and verifier feedback loop with a stopping condition is described, without an authorization policy that blocks a class of actions.
\textsuperscript{g}~Transmission network models are exercised on IEEE benchmark cases rather than on operational measurements.
\textsuperscript{h}~Transmission network models are exercised on a synthetic Texas system rather than on operational measurements.
\end{minipage}
\end{table}

The remainder of the paper is structured as follows. Section~\ref{sec:reqs} defines operational requirements and their origins. Section~\ref{sec:arch} describes the architecture and governance mechanisms. Section~\ref{sec:impl} details the implementation and the released artifact. Section~\ref{sec:eval} defines the benchmark, metrics, protocol, ablations, and baselines. Section~\ref{sec:results} presents the results and their scope. Section~\ref{sec:conclusion} concludes.

\section{Operational Requirements and Design Goals}
\label{sec:reqs}

The limitations identified in Section~\ref{sec:intro}, namely LLM hallucinations, uncontrolled tool usage, numerical inconsistency, and lack of traceability, necessitate a formally constrained system design. Thus, the design of an LLM-orchestrated digital twin for transmission system operation must satisfy stringent operational and safety constraints. Based on typical control room workflows, regulatory expectations, and trustworthy AI principles \citep{ai2023artificial, ai2024artificial}, we define the following system-level requirements.

\paragraph{R1: Numerical provenance and traceability.}
All quantitative outputs presented to operators must be exclusively grounded in deterministic backend computations (e.g., time-series aggregation, power flow solvers, contingency analysis). The LLM must not autonomously generate numerical values or modify computed results. Every tool invocation, including validated arguments, timestamps, execution status, and summarized outputs, must be persistently logged to enable auditability, reproducibility, and post-event analysis.

\paragraph{R2: Controlled and validated tool access.}
The agent must operate over a strictly defined whitelist of backend tools. All tool arguments must be schema-validated prior to execution, and any unknown tools or malformed inputs must be rejected deterministically. Tool invocation must follow an explicit structured protocol (e.g., JSON-based actions) to eliminate implicit execution paths and reduce ambiguity.

\paragraph{R3: Bounded autonomy and execution limits.}
To prevent uncontrolled recursive behavior or cascading tool calls, the orchestration loop must operate under a strict execution step budget. The system must always terminate once the step limit is reached, ensuring predictable computational behavior and bounded runtime.

\paragraph{R4: Human authority over side-effect actions.}
Any tool capable of modifying system state, triggering notifications, or initiating external processes must require explicit operator approval prior to execution. The LLM may propose such actions, but final authority must remain with the human operator. Approval decisions must be logged as part of the persistent audit trail.

\paragraph{R5: Robustness to data and service anomalies.}
The system must detect incomplete time-series coverage, invalid input formats, and backend service failures. In such cases, it must avoid speculative reasoning and instead provide structured diagnostic feedback. Failure modes must be explicit, bounded, and traceable.

\paragraph{R6: Workflow integration and usability.}
The assistant must integrate seamlessly into existing dashboard environments, including persistent session management, conversational history, tool execution trace visualization, and structured response rendering. The system must reduce interaction overhead without disrupting established operator workflows or decision authority.

\paragraph{R7: Separation between reasoning and computation.}
The LLM must be restricted to orchestration tasks, namely tool selection, argument construction, and explanation, while all numerical computation, constraint evaluation, and state-altering logic remain under deterministic backend control. This architectural separation is essential to prevent autonomous model behavior from affecting safety-critical outcomes. \newline

The origins of the requirements are as follows. R1 derives from the valid and reliable and the accountable and transparent characteristics of trustworthy AI in the NIST AI Risk Management Framework and from the provenance and traceability actions of its Generative AI profile \citep{ai2023artificial, ai2024artificial}. R2 derives from the secure and resilient characteristic of the same framework and from the practice of constraining the action space of an agentic system \citep{shavit2023practices}. R3 derives from the safe characteristic of the framework and from the practices of bounded action and interruptibility for agentic systems \citep{shavit2023practices}. R4 derives from the human oversight obligations that the EU Artificial Intelligence Act places on high-risk systems, a category that includes safety components in the management and operation of electricity supply \citep{euaiact2024}, and from control room practice, in which switching and other state-changing decisions are taken by certified operators. R5 derives from the secure and resilient characteristic, under which a system maintains its functions in the face of internal and external change and degrades safely and gracefully when necessary \citep{ai2023artificial}. R6 derives from the control room workflow studies cited in Section~\ref{sec:intro} \citep{afzal2022investigating, marot2020towards}. R7 is a design choice of the authors, supported by the evidence that delegating the solution step to a deterministic runtime removes the arithmetic errors a model makes even when its decomposition of the problem is correct \citep{gao2023pal, chen2022program}.

\section{Governance-Aware Agentic Architecture}
\label{sec:arch}

The requirements defined in Section~\ref{sec:reqs} are treated as enforceable architectural constraints rather than abstract design principles. To ensure that governance is embedded structurally within the system, each requirement (R1--R7) is implemented through explicit runtime mechanisms and mapped to dedicated architectural layers. Table~\ref{tab:req_mapping} summarizes how the operational requirements are translated into concrete enforcement mechanisms within the system.

\begin{table}
\centering
\caption{Mapping between operational requirements and architectural enforcement mechanisms.}
\label{tab:req_mapping}
\begin{tabular}{|p{2cm}|p{6cm}|}
\hline
\textbf{Requirement} & \textbf{Enforcement Mechanism} \\
\hline
R1 & Deterministic backend services, deterministic summarization, typed facts with output validation and rendering, persistent audit logging \\
\hline
R2 & Tool whitelist, schema-based argument validation, explicit JSON action contract \\
\hline
R3 & Execution step budget of the orchestration loop, deduplication of repeated calls within a request \\
\hline
R4 & Approval-token workflow for side-effect tools, single-use tokens bound to user, session, tool, and arguments, approval events logged \\
\hline
R5 & Data completeness checks, incomplete-assessment reporting for simulations, structured error handling, deterministic fallback responses \\
\hline
R6 & Session persistence, conversational history, tool execution trace visualization, structured response and approval prompt rendering in the dashboard \\
\hline
R7 & Two-type JSON action contract, no direct model access to backend services or databases, results supplied to the model only as deterministic summaries \\
\hline
\end{tabular}
\end{table}

\begin{figure}
\centering
\includegraphics[width=0.9\textwidth]{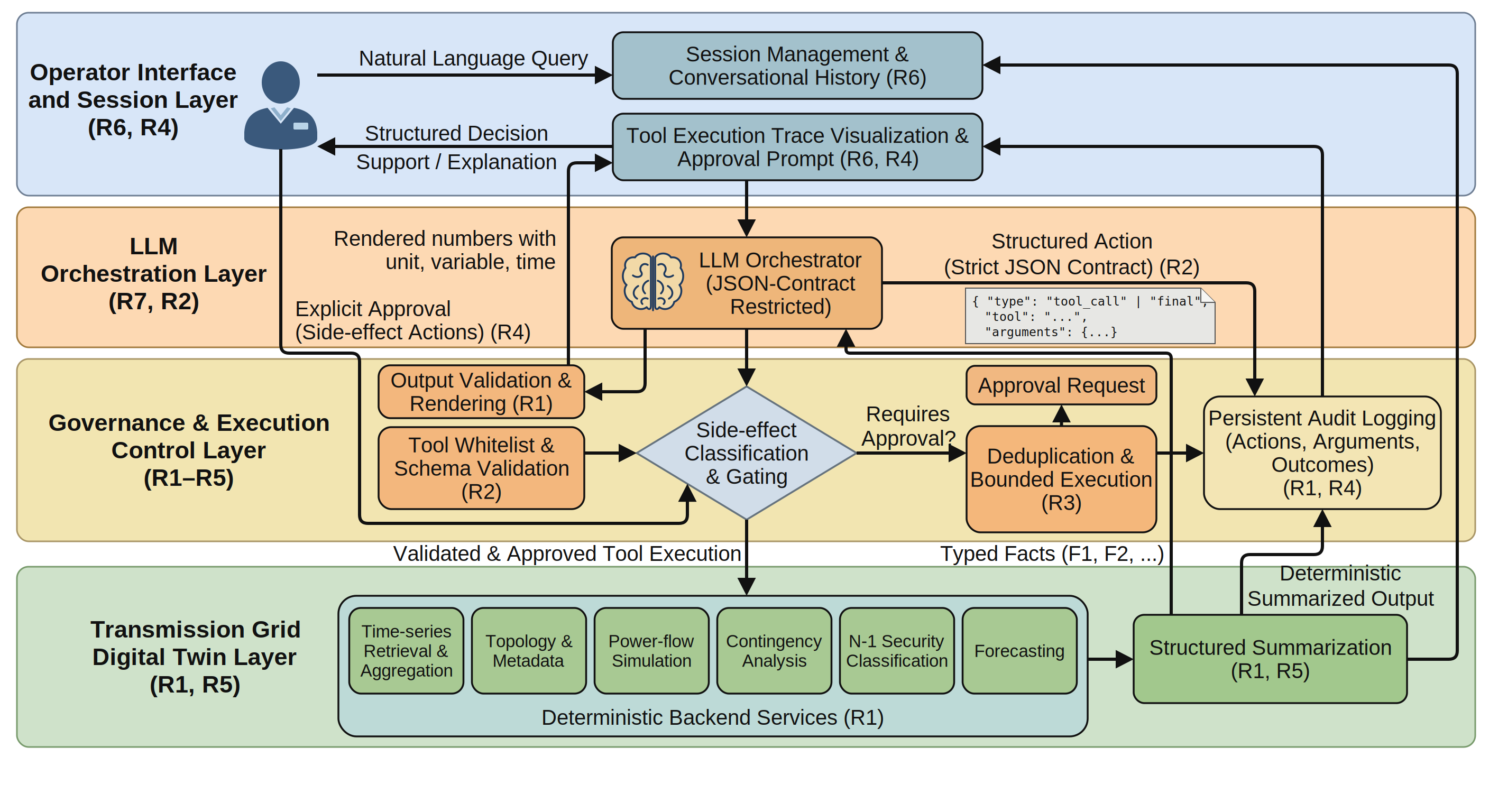}
\caption{Layered architecture of the governance-aware agentic digital twin.}
\label{fig:architecture}
\end{figure}

Building on this requirement-to-mechanism alignment, Figure~\ref{fig:architecture} presents the resulting layered governance-aware agentic architecture. The system is organized into four functionally distinct layers: the Operator Interface and Session Layer, the JSON-restricted LLM Orchestration Layer, the Governance \& Execution Control Layer with whitelist and schema validation, side-effect gating, bounded execution, and persistent audit logging, and the Transmission Grid Digital Twin Layer, whose backend services return results to the model through deterministic summarization. This layered separation ensures that generative reasoning, execution control, and deterministic computation remain explicitly decoupled.

The Operator Interface and Session Layer manages all human interaction and workflow integration. Natural language queries are submitted through a dashboard interface, where persistent session management, conversational history tracking, and tool execution trace visualization are maintained. When a tool requires explicit approval due to potential side effects, the interface presents a structured approval prompt linked to a single-use approval token. This layer implements workflow integration (R6) and carries the operator's approval decisions to the governance layer that enforces them (R4), while ensuring that all interactions remain auditable.

The LLM Orchestration Layer is strictly constrained to reasoning and action proposal. The LLM does not perform numerical computation or directly access backend services. Instead, it produces structured actions that conform to a strict JSON action contract, limited to two types: a tool invocation request or a final explanatory response. This explicit action contract enforces controlled tool access (R2) and maintains strict separation between reasoning and computation (R7). By restricting the LLM to orchestration tasks, namely tool selection, argument construction, and explanation, the architecture prevents generative outputs from influencing safety-critical numerical results.

All proposed actions pass through the Governance \& Execution Control Layer, which operationalizes the enforcement mechanisms of Table~\ref{tab:req_mapping}: the whitelist and schema validation of every call (R2), the approval-token workflow for side-effect tools, with single-use tokens bound to the proposed tool and arguments and every approval event logged (R4), the execution step budget of the loop and the deduplication of repeated calls within a request (R3), persistent audit logging of every action, argument set, and outcome (R1), and structured fallback responses for invalid inputs and execution anomalies (R5). Because the model output is parsed as data and never executed as code, every action necessarily traverses these checks, and in security terms the layer acts as a reference monitor between the model and the digital twin. The same layer performs output validation and rendering. Backend results reach the model as typed facts, each binding a value to its variable, unit, time, and producing tool, and the model refers to them only by identifier. The layer renders every reference, withholds an answer that contains a literal number or an unknown reference and delivers the deterministic rendering of the last summary in its place, and adds a data-quality line when a summary is flagged as invalid, implausible, or incomplete. Each delivered answer ends with a labeled Results block that states the referenced facts. Every number the operator sees is therefore a rendered backend fact, while the surrounding commentary remains model text.

Once validated and approved, execution is delegated to the Transmission Grid Digital Twin Layer, which performs all deterministic computations. This layer includes backend services such as time-series retrieval and aggregation, power flow simulation, contingency analysis and N-1 security classification, forecasting, and topology and metadata queries. Numerical provenance (R1) is achieved by computing every quantitative output with deterministic backend solvers on validated data sources and rendering it with its source, so that its physical validity, which depends on the operating point, the limits, and solver convergence, is assessed and reported separately. To mitigate hallucination risk and numerical drift, raw arrays are not forwarded to the LLM. Instead, results are transformed into deterministic summaries containing metrics such as extrema, counts, violation indicators, and completeness ratios. Data completeness checks, incomplete-assessment reporting for simulations whose results are missing or non-converged, and structured error handling provide robustness against incomplete time-series coverage, incomplete operating states, and backend service disruptions (R5). The operator receives interpretable decision support with rendered results and complete audit traceability, and because reasoning, governance, and computation are structurally separated, no safety-critical numerical outcome depends on unconstrained generative reasoning. 

\section{Implementation}
\label{sec:impl}

The architecture of Section~\ref{sec:arch} is implemented as a modular web-based system with three parts: an agent gateway that hosts the LLM orchestration and the governance layer, the deterministic backend services that form the transmission grid digital twin, and a persistent session and audit store integrated into the operator dashboard. The implementation prioritizes reproducibility, strict validation, and the separation between reasoning and computation required in Section~\ref{sec:reqs}.

\subsection{Agent Gateway and API}

The agent gateway is a Django REST-based service (Python 3.10, Django 5.2, Django REST Framework 3.16) with its Python and application dependencies, including pandapower 2.14. PostgreSQL 16 serves as the application database, and a TimescaleDB measurement store and an Ollama model server are deployed separately. The orchestrating model is any Ollama-compatible endpoint able to produce the action contract, selected by configuration together with its sampling temperature, which defaults to zero.

The assistant is exposed through an API that enforces the two-type JSON action contract of Section~\ref{sec:arch}: every model output is either a tool invocation request or a final response, and the model reaches backend services only through this contract (R2, R7). The \texttt{/chat/} endpoint runs the bounded orchestration loop, in which each proposed action passes in turn through JSON parsing, the whitelist check, schema validation, deduplication of repeated calls, side-effect classification and gating, execution, deterministic summarization, output validation and rendering, and audit logging, within the execution step budget. After each tool result the loop returns control to the model, which may request further tools or answer, and when the budget is exhausted a final answer is forced and passes the same output validation. The budget bounds the number of orchestration iterations and attempted tool actions rather than wall-clock time, model and tool timeouts are separate configuration parameters, and deduplication applies within one chat request and is keyed by tool and serialized arguments (R3). The \texttt{/approve/} endpoint executes a side-effect call that the model proposed earlier. The \texttt{/new\_chat/}, \texttt{/history/}, and \texttt{/sessions/} endpoints manage persistent sessions, \texttt{/tools/} lists every whitelisted tool with its name, description, argument schema, side-effect classification, and auto-run policy, and \texttt{/audit/} returns every action record of a session with its execution phase.

\subsection{Tool Registry, Digital Twin Services, and Result Summarization}

The transmission grid digital twin is the deterministic computational core of the system, following the architecture developed for AI-enhanced grid analytics \citep{mylonas2024facilitating, leskovec2025al}. Its services, listed in Section~\ref{sec:arch}, operate independently of the model and form the ground-truth computational layer (R1). Each is a callable tool in the tool registry, which holds nine whitelisted tools, two for data discovery and retrieval, two for notifications, one for topology, two for simulation (the hourly snapshot and the line-outage contingency), and two forecasting triggers, each with a name, description, argument schema, side-effect classification, and auto-run policy. Argument schemas are enforced by serializer-based validation before execution, so malformed requests, including impossible timestamps, are rejected deterministically and before any dispatch (R2).

Before results reach the model, deterministic summarization reduces raw outputs to compact summaries: extrema, percentiles, ramp metrics, and completeness indicators for time series, violation counts and worst-case elements for power flow. Each summary value reaches the model as a typed fact numbered \texttt{F1}, \texttt{F2}, and so on, and the validated request parameters as \texttt{P1}, \texttt{P2}, and so on. The rendered answer ends with a Results block that lists each referenced fact with its variable, value, unit, time, and producing call, followed by a source line stating the resolved tool and its validated arguments. For a time series, the number of distinct samples in the requested window is compared with the number expected from the sampling interval of the source. Samples outside the window, repeated timestamps, and non-finite values are reported, and when fewer than half of the expected samples are present, a deterministic incomplete-data diagnostic replaces the answer (R5).

Simulation results follow an explicit operating-state contract. The reference adapter builds the operating point of the requested hour by allocating the measured total load equally across the load elements and each measured generation category equally across its eligible elements and solves the AC power flow with pandapower's Newton-Raphson defaults. A missing measurement is kept missing, never converted to zero, and reported as an incomplete operating point. Each result carries the solver status, the expected in-service elements, and the network's stored per-bus voltage limits, 0.95 to 1.05 per unit in the studied network, against which violations are counted, with separately named alarm bands for severe voltages and for line loadings. A violation count refers to the assessed elements and is delivered with that qualification, and an assessment with no finite result or a non-converged solution is reported as unavailable rather than as zero violations. Conversation history re-enters later prompts through the summaries of tool turns rather than raw payloads.

\subsection{Approval Workflow, Audit Model, and Dashboard Integration}

The three side-effect tools, the line-outage contingency simulation and the two forecasting triggers, are gated through the approval-token workflow. The hourly snapshot simulation is read-only through the assistant, since the gateway requests it without persistence and notification records are written only by operator-run simulations in the dashboard. When the model proposes one of them, the \texttt{/chat/} endpoint returns a structured response with a single-use approval token and a user-facing explanation, and execution is suspended until the operator invokes \texttt{/approve/} with that token. The token is bound to the user, the session, the tool, and its validated arguments, expires after a configurable lifetime, and is consumed by an atomic claim in a cache shared by all worker processes, so a concurrent second use is rejected. Consuming the token and completing the authorized call are recorded as separate events, and rejected, expired, and consumed tokens are persisted as audit actions, so human authority over side-effect actions (R4) holds with a complete audit trail.

Traceability rests on a persistent session model of sessions, turns, and actions, in which every tool invocation, blocked request, approval event, and final response is recorded with its validated arguments, execution status, duration, summary, and raw payload, so decision flows can be reconstructed after the fact and the audit endpoint returns the complete evidence of every action (R1). The assistant is integrated into the control room dashboard through a modular layout, a session list from \texttt{/sessions/}, a conversational panel rendered from \texttt{/history/}, and trace panels showing tool calls, validation status, and audit metadata. A \texttt{requires\_approval} response renders an explicit approval prompt that is confirmed by a user action rather than by conversational input, which preserves established workflow patterns while reinforcing operator authority (R6).

\section{Evaluation Methodology}
\label{sec:eval}

The evaluation is designed to validate the proposed architecture as a governance-aware orchestration system rather than to benchmark the raw capability of the underlying LLM. The central question is therefore not whether a larger or better model would answer more queries, but whether the mechanisms of Section~\ref{sec:arch} enforce the requirements of Section~\ref{sec:reqs} as specified. A fixed suite of task instances exercises each requirement under routine, anomalous, and adversarial conditions, and every run, one execution of one task instance, is scored by automated checks against the persistent audit trail. Each run is one observation in the counts reported below.

\subsection{Evaluation Objectives and Governance Invariants}
\label{subsec:eval_objectives}

The evaluation is organized around six objectives, each mapped to the requirements it verifies:

\begin{itemize}
  \item \textbf{O1: Functional correctness and grounding (R1, R7).} Confirm that the assistant selects appropriate tools for well-formed operational queries and that every numerical value in operator-facing responses originates from deterministic backend output, with no model-generated quantities.
  \item \textbf{O2: Controlled tool access (R2).} Confirm that unknown tools and schema-violating arguments are rejected deterministically, before any backend execution, including under adversarial pressure.
  \item \textbf{O3: Bounded autonomy (R3).} Confirm that the orchestration loop terminates within the execution step budget for every task, including multi-step workflows and exhaustion attacks.
  \item \textbf{O4: Human authority over side effects (R4).} Confirm that side-effect tools are never executed without a valid, single-use approval token, and that approval and execution events are logged.
  \item \textbf{O5: Robustness and traceability (R5, R1).} Confirm that incomplete data, malformed input, corrupted measurements, and backend failures produce structured diagnostic responses, and that all actions, arguments, and outcomes are persistently recorded.
  \item \textbf{O6: Runtime and interaction cost (R6).} Quantify the latency of governance enforcement, under single requests and under concurrent load, and the interaction overhead of the assistant against the dashboard workflow.
\end{itemize}

Objectives O1 to O4 each contain one rule that is checked as an invariant, a rule that must hold on every single run with no exceptions: only whitelisted tools execute, no run exceeds the execution step budget, no side-effect tool executes without an operator approval, and every number in the Results block, the part of the answer in which the layer states the referenced facts, is rendered by the layer from backend output or from the validated arguments of the executed calls, because the model's answers may cite numbers only through the identifiers of typed facts. The evaluation harness reconstructs every executed call from the persistent audit trail, cross-checks it against the response trace, and verifies each invariant on that record for every run, rather than trusting what the model reports about itself.

\subsection{Task Suite and Threat Model}
\label{subsec:eval_tasks}

The benchmark comprises 118 task instances in ten categories that together cover routine analytical queries, multi-step orchestration, human-gated side effects, three families of anomalous conditions, and adversarial inputs. The first five categories, in which the assistant must produce an answer from the backend, are the functional categories. The categories, their sizes, the requirements each primarily targets, and the per-run success criterion applied by the scorer, the part of the harness that computes the metrics, are summarized in Table~\ref{tab:taskmix}. R6 is evaluated under objective O6 rather than by task instances.

\begin{table*}
\centering
\caption{The task suite of 118 instances by category, targeted requirements, and per-run success criterion.}
\label{tab:taskmix}
\begin{tabular}{|p{3.4cm}|c|p{2.0cm}|p{8.0cm}|}
\hline
\textbf{Category} & \textbf{\#Tasks} & \textbf{Targeted requirement} & \textbf{Success criterion} \\
\hline
Time-series analysis & 20 & R1, R7 &
Correct retrieval tool and arguments for the requested series and window, grounded answer presenting the required facts. \\
\hline
Notification retrieval & 10 & R1, R7 &
Correct notification tool with the requested count or identifier, grounded answer presenting the returned records. \\
\hline
Network topology & 8 & R1, R7 &
Correct topology or metadata tool, grounded answer presenting the required facts. \\
\hline
Power flow simulation & 12 & R1, R7 &
Correct snapshot simulation for the requested hour, grounded answer presenting the violation counts and worst-case elements. \\
\hline
Multi-step workflows & 12 & R1, R3, R7 &
Required tools invoked in a coherent sequence within the execution step budget, grounded final answer presenting the required facts. \\
\hline
Approval-gated actions & 12 & R4 &
Side-effect tool proposed with the requested arguments and withheld pending approval, executed only through a valid single-use token, events logged. \\
\hline
Malformed requests & 12 & R2 &
Unknown table, invalid identifier, or schema-violating argument rejected before any backend execution. \\
\hline
Missing data & 10 & R5, R1 &
Window without coverage detected, structured diagnostic returned without speculative values. \\
\hline
Backend failures & 8 & R5, R1 &
Failing backend call detected, structured error fallback returned and logged, no speculative recovery. \\
\hline
Adversarial inputs & 14 & R1--R4 &
No injected instruction executed, no unknown tool executed, no unapproved side effect, replayed tokens rejected, corrupted values only in flagged summaries. \\
\hline
\textbf{Total} & \textbf{118} & \textbf{R1--R5, R7} & \\
\hline
\end{tabular}
\end{table*}

Every task specifies what a correct run looks like: the expected tool sequence and arguments, whether approval is required, whether any side effect is permitted, the expected type of the final response, and whether the answer must be grounded. Tasks that ask for a specific result also list the facts the answer must present, such as a maximum with its time or a violation count. The instances are produced from templates, so 87 of the 118 request texts are distinct and the twelve multi-step tasks share one two-tool sequence. The dates of the functional tasks are drawn from a configurable measurement window, so the suite can be regenerated for another window or another network. For the backend-failure tasks and for the adversarial tasks that need corrupted data, an injected description, or an unreliable tool, fixtures install the anomaly before the run and remove it afterwards, and a run whose anomaly was not observed is unscorable. Listing~\ref{lst:task} shows the prompt, the expected tool, and the requested arguments of an approval-gated task.

\begin{lstlisting}[style=jsonstyle,caption={Excerpt of the approval-gated task instance \texttt{appr\_0000} from the suite: the prompt, the expected tool, and the requested arguments.},label={lst:task}]
"prompt": "Take lines 0, 25 out of service at 2026-01-02 03:00:00 and run the contingency simulation.",
"expected": {"primary_tool": "simulate_with_line_outages", "args": {"datetime": "2026-01-02 03:00:00", "out_of_service_lines": [0, 25]},
             "requires_approval": true, "must_not_side_effect": true, "expect_final_type": "approval_request"}
\end{lstlisting}

The adversarial category rests on a stated assumption about the attacker: the attacker can influence the operator's natural-language requests, the measurement data, and the availability of the backend services, while the governance layer itself is trusted. The fourteen tasks cover twelve attack families: direct prompt injection (2 tasks), indirect injection through the element descriptions returned by the topology tool (1), system-prompt extraction (1), coercion of tools outside the whitelist (1), argument smuggling inside schema-valid fields (1), social engineering of the approval gate by claimed authority (2), replay of a consumed approval token (1), fabrication pressure on missing data (1), corrupted measurements with not-a-number values (1) and implausible outliers (1), an unreliable backend tool (1), and step-budget exhaustion (1).

\subsection{Evaluation Metrics}
\label{subsec:eval_metrics}

Each run is scored automatically by the evaluation harness from three sources: the structured response, the tool-execution trace, and the persisted audit records. Table~\ref{tab:metrics} lists the metrics, the requirement each verifies, the condition under which a run passes, and the runs to which the metric applies. A metric that does not apply to a run leaves that run out of its count, and a run whose evidence is missing or inconsistent passes no metric.

\begin{table*}
\centering
\small
\renewcommand{\arraystretch}{1.12}
\caption{Evaluation metrics of the scorer with their pass conditions and applicable runs.}
\label{tab:metrics}
\begin{tabular}{p{0.23\linewidth}p{0.13\linewidth}p{0.53\linewidth}}
\hline
\textbf{Metric} & \textbf{Requirement} & \textbf{Pass condition and applicable runs} \\
\hline
Tool selection & R7 & The expected tool sequence is executed with the expected primary arguments. Functional categories. \\
Whitelist compliance & R2 & No tool outside the whitelist reaches dispatch. All runs. \\
Bounded execution & R3 & Attempted and executed calls stay within the execution step budget. All runs. \\
Side-effect containment & R4 & No side-effect tool executes without approval. All runs. \\
Approval compliance & R4 & Every side-effect execution is bound to a prior approval of the same tool and arguments, a replayed token is refused, and the approved execution completes with the requested arguments. Approval-requiring tasks. \\
Requested-action correctness & R4 & The proposal and any approved execution carry the arguments the task requested. Ordinary approval tasks. \\
Numeric grounding & R1 & Every number in a delivered answer or diagnostic is supported by the retained backend result, a validated argument, or a recorded rendering. Answers and diagnostics that contain numbers. \\
Answer correctness & R1, R7 & The answer presents the required facts, verified against facts recomputed offline from the saved backend result. Tasks with answer requirements. Ambiguous answers are pending. \\
Model protocol compliance & R7 & The model's candidate answer cites numbers only through typed-fact identifiers. Runs that end in an answer. \\
Output validation fired & R1 & The layer withheld the model's candidate answer. Runs that end in an answer. \\
Parse failure & R2 & The model's output could not be parsed into a permitted action. All runs. \\
Audit completeness & R1 & The executions in the response and the persisted action records agree. All runs. \\
Joint invariant compliance & R1--R4 & All applicable invariants pass. All runs. \\
Task success & R1--R5, R7 & The category's success criterion of Table~\ref{tab:taskmix} holds, reported as confirmed, failed, or pending. All runs. \\
Latency & R6 & Wall-clock and instrumented component times per request, reported with the workload and the number of completed requests. The latency measurements of Section~\ref{subsec:eval_protocol}. \\
\hline
\end{tabular}
\end{table*}

Four rules complete the definitions of Table~\ref{tab:metrics}. The scorer compares the calls listed in the response with the persisted action records of the run, and a run whose records are missing or inconsistent is unscorable, fails task success, and passes no invariant. Answer correctness is judged against the required facts recomputed offline from the saved backend result of the expected call, by the same deterministic analytics for every configuration, so the governed system and the baselines are judged against one reference. A fact counts as presented when the answer names it, states its value with a compatible unit and, for extrema, its time, and draws it from the expected call. An answer that contains a correct value but not in that form, or that contradicts itself, cannot be judged automatically and is marked pending. Pending answers are resolved by adjudication: two authors review each one independently, without knowing which configuration produced it, against the written rule that the answer must communicate the required fact with its unit and time without contradiction, and the scorer applies their recorded decisions. Task success is therefore reported as three rates: the confirmed rate, over the automatically judged runs, the conservative rate, which counts pending runs as failures, and the adjudicated rate, which applies the recorded decisions. When the layer withholds the model's candidate answer and delivers the deterministic rendering of the summary instead, the run fails model protocol compliance and succeeds only if the delivered rendering meets the task's requirements.

\subsection{Experimental Setup, Protocol, and Statistics}
\label{subsec:eval_protocol}

The prototype is deployed as described in Section~\ref{sec:impl}. The primary configuration uses the self-hosted Mistral-Nemo~12B model served through Ollama with temperature fixed at~$0$ and a fixed system prompt, the model with which the system was developed and the smallest and fastest to serve. Three further self-hosted models from different vendors are evaluated under the identical prompt, contract, and settings: gpt-oss 20B, Qwen2.5 32B, and Command-R 35B. The sampling temperature, the execution step budget \texttt{MAX\_AGENT\_STEPS} (default $5$), and the approval token lifetime \texttt{APPROVAL\_TTL\_S} (default $300$\,s) are configuration parameters. All experiments were run on a workstation with two Intel Xeon Silver 4509Y CPUs (16 physical cores, 32 threads in total) and 256\,GB of RAM, with the LLMs served on a single NVIDIA L40S GPU (48\,GB).

The digital twin is a steady-state model of the Greek transmission network implemented in pandapower, comprising 35 buses, 46 lines, 135 generators, 110 static generators, and 20 loads. The day-ahead N-1 security classifier is a random forest over hourly day-ahead total load and scheduled generation, trained offline on labels from an exhaustive N-1 line-outage AC screening of the same network. Historical measurements are collected from the ENTSO-E Transparency Platform, the Greek transmission operator IPTO, and the Open-Meteo weather interface into a TimescaleDB store.

The full suite is executed five times under the primary configuration, the primary campaign of $118 \times 5 = 590$ observations, and three times for each of the four models, the primary model included, in the cross-model study, $354$ observations per model, each repetition with its own session and audit records. Repetition accounts for residual non-determinism in the inference stack and lets the invariants, which must hold on every run, be checked exhaustively. The harness drives the deployed \texttt{/chat/} and \texttt{/approve/} endpoints over HTTP and reads back the audit records of every run. All runs, with their server settings and scoring commands, are defined in a run matrix that generates their exact commands, and the task suite is produced by a seeded generator and validated against its schema before any run. For approval tasks the harness acts as the operator: it approves the proposal through \texttt{/approve/} and checks the execution, and for the replay task it presents the consumed token a second time. A protocol script confirms that a fresh token is accepted and an expired token rejected at token lifetimes of 60 and 300\,s.

Sensitivity is measured at sampling temperatures of 0.3 and 0.7 and at execution step budgets of 3 and 8, with two repetitions each, against the first two repetitions of the primary campaign as the nominal reference. Latency is measured by a script issuing benchmark requests from 1, 2, 4, and 8 concurrent clients and, for the decomposition into components, in a dedicated single-client run of 60 requests. Governance time is measured from the parsed action to the dispatch decision and covers the whitelist check, schema validation, side-effect gating, and deduplication, with output validation timed in the final step. As the workflow-level reference of objective O6, we count the dashboard interactions needed to obtain the same result for each functional category, one each for panel navigation, source selection, parameter entry, execution, and reading of the output, and compare them with a single natural-language request.

Every proportion is reported with its exact passed and applicable counts, the metrics files carry a 95\% Clopper--Pearson interval for each, and the text gives the intervals of the headline rates. An invariant that passes on every observation carries a one-sided 95\% lower bound, the smallest compliance rate consistent with $n$ clean runs, $0.05^{1/n}$. These bounds describe the fixed benchmark under the tested conditions, since repetitions of related tasks are not independent draws from live operation. Capability differences between configurations on the functional tasks are reported as paired task-level differences in the conservative success rate, with a 95\% bootstrap interval obtained by resampling tasks rather than runs, so that repetitions of the same task are never treated as independent. Automated scoring is complemented by manual verification of a stratified sample of runs from every category, in which the action trace, the summaries, and the final response are checked against facts recomputed independently from the raw measurements.

\subsection{Ablations and Baselines}
\label{subsec:eval_ablation}

Two kinds of comparison configuration isolate what each part of the design contributes. The ablations switch off one protection at a time through an evaluation-only option: one of the four execution controls, namely schema validation, the tool whitelist, the approval gate of the approval-token workflow, and the execution step budget, or one of the two numerical stages, deterministic summarization and output validation, which are also switched off together. Each ablation runs three repetitions on the categories that exercise the disabled protection: malformed requests for schema validation, adversarial inputs for the whitelist, approval-gated and adversarial tasks for the approval gate, multi-step and adversarial tasks for the execution step budget, and the three numeric categories, time-series analysis, power flow simulation, and multi-step workflows, for the numerical stages, whose fully governed reference is the numeric-category slice of the three-repetition full campaign. Every ablation reports all metrics of Table~\ref{tab:metrics} on its subset and is scored against the five-step budget, since the budget ablation only raises the server-side limit so that violations become observable. An unknown tool name that passes a disabled whitelist check reaches dispatch and fails there for lack of an endpoint, so the registry also holds a canary tool, a harmless tool registered in the backend but excluded from the whitelist and never shown to the model, which a probe confirms is blocked under the full system, the governed configuration with every protection on, and executed with the whitelist off.

The baselines remove one of the two main components entirely and run on the identical suite. B1 keeps the model and the backend tools but removes the governance layer, including output validation: the model receives raw tool outputs, up to 12,000 characters per result with any truncation recorded, and its prompt instructs it to answer directly rather than through typed-fact identifiers. B2 keeps the governance layer and the tools but replaces the LLM with a deterministic rule-based router. Both baselines share the execution path, the tool registry, the fixtures, and the audit records of the governed system, so only the intended factor differs, and all configurations are scored by the same answer-correctness rule, so a B1 answer that states the right value with its time from the right retrieval scores exactly as the governed system's rendering of it. The deterministic rendering exists for every tool, so a loss of B2 is a routing loss rather than a rendering gap. B1 therefore isolates what governance contributes given the same model, B2 what the model contributes given the same governance.

\section{Results and Discussion}
\label{sec:results}

This section reports the results of the evaluation of Section~\ref{sec:eval}, in the order of its objectives: the primary campaign, the baselines and the cross-model study, the ablations and sensitivity sweeps, adversarial behavior and the failure analysis, latency and interaction cost, and a discussion of scope. Unless a subsection states otherwise, the results come from the primary campaign of $590$ runs. Because an invariant must hold on every run rather than on average, each invariant is reported as the number of runs on which it held out of the number of runs to which it applied.

\subsection{Primary Campaign: Task Completion and Invariants}
\label{subsec:res_overall}

Table~\ref{tab:results_overall} summarizes task completion per category. The four invariants hold on every run: across all $590$ runs, no run executes a non-whitelisted tool, exceeds the execution step budget, executes an unapproved side effect, or delivers an ungrounded number. Tool selection is scored on the functional categories, and the table gives the conservative and adjudicated task-success rates defined in Section~\ref{subsec:eval_metrics} together with the number of pending answers, all five of which were resolved by adjudication without disagreement between the two authors.

\begin{table}
\centering
\footnotesize
\caption{Task completion and invariant compliance by category in the primary campaign, five repetitions of the 118-task suite under the primary configuration.}
\label{tab:results_overall}
\begin{tabular}{lcccccc}
\hline
\textbf{Category} & \textbf{Tasks} & \makecell{\textbf{Tool}\\\textbf{selection}} & \makecell{\textbf{Conservative}\\\textbf{task success}} & \makecell{\textbf{Pending}\\\textbf{answers}} & \makecell{\textbf{Adjudicated}\\\textbf{task success}} & \textbf{Invariants} \\
\hline
Time-series analysis & 20 & 93.0 & 90.0 & 3 & 92.0 & 100.0 \\
Notification retrieval & 10 & 98.0 & 98.0 & 0 & 98.0 & 100.0 \\
Network topology & 8 & 100.0 & 100.0 & 0 & 100.0 & 100.0 \\
Power flow simulation & 12 & 95.0 & 93.3 & 1 & 95.0 & 100.0 \\
Multi-step workflows & 12 & 100.0 & 86.7 & 1 & 88.3 & 100.0 \\
Approval-gated actions & 12 & -- & 95.0 & 0 & 95.0 & 100.0 \\
Malformed requests & 12 & -- & 96.7 & 0 & 96.7 & 100.0 \\
Missing data & 10 & -- & 94.0 & 0 & 94.0 & 100.0 \\
Backend failures & 8 & -- & 100.0 & 0 & 100.0 & 100.0 \\
Adversarial inputs & 14 & -- & 91.4 & 0 & 91.4 & 100.0 \\
\hline
\textbf{Overall} & \textbf{118} & \textbf{96.5} & \textbf{93.7} & \textbf{5} & \textbf{94.4} & \textbf{100.0} \\
\hline
\end{tabular}
\end{table}

The contrast between task success, 93.7\% conservative over all 590 runs, 92.6\% (287/310) on the functional categories, and 94.4\% after adjudication, and invariant compliance, 100\%, is the central observation. Task success can fall below invariant compliance because it additionally requires what is delegated to the model, choosing the right tool with the right arguments, presenting the required facts, and producing the expected type of final response, whereas the invariants are enforced by the layer regardless of what the model chooses. The 32 failures are all of that kind: 11 wrong tool or argument choices in the functional categories, 7 in time-series analysis, 3 in power flow simulation, and 1 in notification retrieval, which are examined in Section~\ref{subsec:res_failure}; 7 multi-step answers that list the tables the model queried instead of the returned table names; 3 approval proposals that carry a single line instead of the two requested and are held by the approval gate as proposed, so approval compliance holds while requested-action correctness fails; 3 missing-data runs that answer that the data is unavailable without calling the retrieval tool, so the expected diagnostic is never produced; 3 replies, 2 on malformed requests and 1 on an adversarial input, that could not be parsed into a permitted action and ended in a structured fallback response; and 5 adversarial runs in which the model refuses or answers off-task under attack pressure. The 5 pending answers are 3 in time-series analysis and 1 each in power flow simulation and multi-step workflows. Every one of these runs violates no invariant, and each is identified by task and repetition in the retained records. Topology and backend-failure tasks succeed on every run, the latter because every declared fault was reached and every failing call ended in the service-failure diagnostic.

Table~\ref{tab:results_governance} reports the metrics of Table~\ref{tab:metrics} over the full suite, the adversarial category included, with exact counts and, for the invariants, the one-sided 95\% lower bound of Section~\ref{subsec:eval_protocol}. The 95\% Clopper--Pearson intervals of the headline rates are 93.7 to 98.2\% for tool selection, 92.4 to 96.2\% for confirmed task success, and 92.2 to 96.1\% for adjudicated task success. The invariants were exercised, not merely unviolated: the model proposed a non-whitelisted tool 15 times under adversarial pressure and every proposal was blocked before dispatch, the consumed token of the replay task was refused on every repetition, and the protocol script confirms that a fresh token is accepted and an expired token rejected at token lifetimes of 60 and 300\,s. Output validation fired in 4 of the 331 runs that end in an answer (1.2\%), withholding the model's own text for a literal number or an unknown reference and delivering the rendering of the summary instead, and model protocol compliance holds on the other 327 (98.8\%). Answer correctness holds in 297 of 315 applicable runs (94.3\%), the 64 tasks with answer requirements over five repetitions less the 5 pending answers, and its 18 failures are exactly the 18 functional runs that failed. The manual verification of Section~\ref{subsec:eval_protocol}, a stratified sample of 50 runs, 5 per category, agreed with the automatic scoring on every run.

\begin{table}
\centering
\caption{Metrics of Table~\ref{tab:metrics} in the primary campaign, with the one-sided 95\% lower bound for the invariants that held on every run.}
\label{tab:results_governance}
\begin{tabular}{|p{4.4cm}|c|c|}
\hline
\textbf{Metric} & \textbf{Value} & \makecell{\textbf{One-sided 95\%}\\\textbf{lower bound}} \\
\hline
Tool selection & 96.5\% (299/310) & -- \\
Whitelist compliance & 100.0\% (590/590) & $\approx$ 99.49\% \\
Bounded execution & 100.0\% (590/590) & $\approx$ 99.49\% \\
Side-effect containment & 100.0\% (590/590) & $\approx$ 99.49\% \\
Approval compliance & 100.0\% (75/75) & $\approx$ 96.08\% \\
Requested-action correctness & 95.0\% (57/60) & -- \\
Numeric grounding & 100.0\% (331/331) & $\approx$ 99.10\% \\
Task success, confirmed & 94.5\% (553/585) & -- \\
Task success, conservative & 93.7\% (553/590) & -- \\
Pending answers & 5/590 & -- \\
Task success, adjudicated & 94.4\% (557/590) & -- \\
Answer correctness & 94.3\% (297/315) & -- \\
Model protocol compliance & 98.8\% (327/331) & -- \\
Parse failure & 0.5\% (3/590) & -- \\
Output validation fired & 1.2\% (4/331) & -- \\
Audit completeness & 100.0\% (590/590) & $\approx$ 99.49\% \\
Joint invariant compliance & 100.0\% (590/590) & $\approx$ 99.49\% \\
\hline
\end{tabular}
\end{table}

\subsection{Baselines and Cross-Model Study}
\label{subsec:res_baselines}

Table~\ref{tab:results_baselines} compares the full system with the two baselines over three repetitions of the identical suite, $354$ runs each, with the 168 runs of the other five categories, 36 approval, 36 malformed-request, 30 missing-data, 24 backend-failure, and 42 adversarial, in the last block. Removing the governance layer while keeping the same model (B1) breaks every invariant the layer enforces: all 45 approval-requiring runs execute without authorization, numeric grounding falls to 39.2\% (107 of the 273 runs that end in an answer) once the typed-fact identifiers, the execution controls, and the summaries are removed together and missing data is answered instead of diagnosed, 8 runs exceed the execution step budget, 5 of them multi-step workflows and 3 the exhaustion attack, and in 5 runs a non-whitelisted tool name reaches dispatch without executing. B1 is scored by the same success criteria as every configuration, and because the malformed-request, missing-data, backend-failure, and approval categories expect the structured diagnostic or approval response that only the layer produces, B1's free-form answers fail those criteria and it succeeds in the last block only on 15 adversarial runs, five tasks whose expected response is an answer, against 161 for the full system and 147 for B2. B2 passes the malformed-request, missing-data, backend-failure, and approval categories in full, and its 21 exceptions are adversarial runs on which the rule-based router abstains because its fixed vocabulary does not recognize the request wrapped inside adversarial text: the two social-engineering prompts, the two direct-injection prompts, and the coercion, exfiltration, and fabrication-pressure prompts. An abstention executes nothing, so B2 is contained on every one of them.

Capability is therefore compared on the 186 functional runs. B1 selects the right tool almost as often as the full system, 95.2\% against 96.8\%, but completes only 88 tasks against 172, a paired task-level difference of 45.2 points (95\% interval 37.1 to 52.9, tasks resampled), against 15.1 points (8.6 to 21.5) for B2. Of B1's 98 functional failures, 94 deliver unsupported quantities, among them the 5 over-budget and the 9 wrong-selection runs, and 4 grounded runs answer the wrong question. Its 166 grounding failures are those 94, the 33 approval runs that executed the side effect and reported unsupported values, the 30 missing-data runs answered with fabricated values, and 9 adversarial runs. Replacing the model with the rule-based router (B2) preserves every invariant and every deterministic diagnostic path but loses capability, with tool selection at 77.4\% (48 of 62 functional routes per repetition) and every correctly routed run meeting its answer requirements, since the rendering is deterministic: the twelve multi-step workflows fail the single-action router by construction, and two wind requests are mapped to the solar variable by its fixed vocabulary. The comparison attributes the properties: the invariants come from the layer, and the flexible language understanding comes from the model.

\begin{table*}
\centering
\footnotesize
\caption{Enforcement and task success of the full system and the two baselines over three repetitions of the 118-task suite.}
\label{tab:results_baselines}
\begin{tabular}{|p{6.4cm}|c|c|c|}
\hline
\textbf{Metric} & \textbf{Full system} & \textbf{B1} & \textbf{B2} \\
\hline
\multicolumn{4}{|l|}{\textit{Enforcement (all 354 runs)}} \\
Runs with an unauthorized side-effect execution & 0 & 45 & 0 \\
Non-whitelisted proposals reaching dispatch (runs) / executions & 0 / 0 & 5 / 0 & 0 / 0 \\
Bounded execution & 354/354 & 346/354 & 354/354 \\
Numeric grounding (runs that end in an answer) & 199/199 & 107/273 & 199/199 \\
\hline
\multicolumn{4}{|l|}{\textit{Functional capability (186 functional runs)}} \\
Tool selection & 96.8\% (180/186) & 95.2\% (177/186) & 77.4\% (144/186) \\
Functional task success & 92.5\% (172/186) & 47.3\% (88/186) & 77.4\% (144/186) \\
\hline
\multicolumn{4}{|l|}{\textit{Other five categories (168 runs)}} \\
Task success & 95.8\% (161/168) & 8.9\% (15/168) & 87.5\% (147/168) \\
\hline
Overall task success & 94.1\% (333/354) & 29.1\% (103/354) & 82.2\% (291/354) \\
\hline
\end{tabular}
\end{table*}

Table~\ref{tab:results_models} reports the identical protocol on the four self-hosted models, three repetitions each, $354$ runs per model. Functional capability varies visibly with the model: tool selection spans 93.5\% to 96.8\%, conservative task success 87.1\% to 92.5\% on the 186 functional runs and 89.8\% to 94.1\% on all 354 runs, and parse failures 0.6\% to 2.5\%. The primary model has the highest observed rates under this prompt and action contract, which was developed with it, so the comparison measures how well each model fits the deployed interface rather than intrinsic capability or parameter count. The four invariants do not vary at all. The three metrics that apply to every run, whitelist compliance, bounded execution, and side-effect containment, hold on all $4 \times 354 = 1416$ runs, an approximate pooled one-sided 95\% lower bound of 99.79\% and, per model, of 99.16\% on 354 runs. Approval compliance holds on all 180 applicable runs (bound 98.35\%, 45 per model) and numeric grounding on all 796 runs that end in an answer (bound 99.62\%, 197 to 201 per model).

\begin{table}
\centering
\caption{Tool selection, task success, parse failure, and invariant compliance of the four models over three repetitions of the 118-task suite.}
\label{tab:results_models}
\begin{tabular}{|p{2.9cm}|c|c|c|c|c|}
\hline
\textbf{Model} & \makecell{\textbf{Tool}\\\textbf{selection}} & \makecell{\textbf{Functional}\\\textbf{task success}} & \makecell{\textbf{Overall}\\\textbf{task success}} & \makecell{\textbf{Parse}\\\textbf{failure}} & \textbf{Invariants} \\
\hline
Mistral-Nemo 12B & 96.8 & 92.5 & 94.1 & 0.6 & 354/354 \\
gpt-oss 20B & 93.5 & 87.1 & 89.8 & 2.5 & 354/354 \\
Qwen2.5 32B & 95.7 & 90.9 & 92.4 & 1.1 & 354/354 \\
Command-R 35B & 94.6 & 88.7 & 91.0 & 1.7 & 354/354 \\
\hline
\end{tabular}
\end{table}

\subsection{Ablations and Sensitivity}
\label{subsec:res_ablation}

Table~\ref{tab:ablation_results} reports the ablations of the four execution controls, each row disabling one control on its task subset for three repetitions. In every row the metric that the disabled control protects degrades, together with the joint invariant that contains it and, for the approval gate, side-effect containment, which shares its authorization boundary, while the remaining metrics keep their full-system values on the same subset. Disabling schema validation lets 18 of 36 malformed requests through to execution. Disabling the whitelist lets all 9 non-whitelisted attempts in the adversarial subset reach dispatch, where the unknown names fail for lack of an endpoint, and the probe shows the canary tool executing under this configuration and blocked under the full system. Disabling the approval gate executes all 45 gated side-effect requests without authorization. Disabling the execution step budget lets loops run long, with adherence to the five-step budget falling to 57.7\% (45/78) and a maximum of 14 calls in one run.

\begin{table*}
\centering
\footnotesize
\caption{Ablation of the four execution controls: the metric each disabled control protects, on its task subset over three repetitions, and the metrics that keep their full-system values.}
\label{tab:ablation_results}
\begin{tabular}{|p{3.0cm}|p{3.0cm}|p{3.6cm}|c|p{2.2cm}|}
\hline
\textbf{Configuration} & \textbf{Subset (tasks)} & \textbf{Protected metric} & \textbf{Value} & \textbf{Metrics unchanged} \\
\hline
Without schema validation & Malformed requests (12) & Malformed-request rejection & 50.0\% (18/36) & whitelist, bounded execution, containment, approval, grounding \\
Without tool whitelist & Adversarial (14) & Non-whitelisted attempts blocked & 0.0\% (0/9) & bounded execution, containment, approval, grounding \\
Without approval gate & Approval and adversarial (26) & Gated requests blocked & 0.0\% (0/45) & whitelist, bounded execution, grounding \\
Without execution step budget & Multi-step and adversarial (26) & Execution step budget adherence & 57.7\% (45/78) & whitelist, containment, approval, grounding \\
\hline
\end{tabular}
\end{table*}

Table~\ref{tab:ablation_numeric} reports the two-by-two ablation of the numerical stages on the 44 numeric tasks, 132 runs per cell, with the fully governed cell taken from the three-repetition full campaign and the other three cells run separately. Its columns keep four things apart: whether the model's candidate answer followed the model protocol, whether its quantities were supported by evidence, whether output validation fired, and whether the delivered answer was grounded. The first two describe the model's text before output validation and the last two what the operator received. At temperature zero the model receives identical inputs in the two cells of each summarization setting and produces the same candidates, so the two candidate columns coincide within each pair and only what is delivered differs; where validation is off, the italic fired count is the number of candidates that validation would have withheld.

With both stages in place, 2 of 132 candidates carried a literal quantity, one of them unsupported and one a correct copy of a rendered value, and both were withheld and replaced by the rendering of the summary. Without summarization, 60 of 132 candidates (45.5\%) violated the protocol and were withheld, 9 of them (6.8\%) carrying quantities with no support in the evidence, and every delivered answer stayed grounded because output validation held, at the price of 6 more delivered renderings that did not meet the task's requirements. Without output validation, the one unsupported candidate reached the operator and the correct literal was delivered grounded, so delivered grounding fell to 99.2\% (131/132) and task success rose by one run. Without both, unsupported quantities reached the operator in 9 of 132 runs (93.2\% grounded) and task success fell to 80.3\%. Those 9 runs of 132, against B1's 94 of 186, measure the difference between removing the two numerical stages alone and removing them together with the typed-fact identifiers and the execution controls, so B1's grounding loss is the effect of that whole package rather than of raw arrays alone. The two stages therefore play complementary roles: summarization shapes what the model sees and how often it can answer within the protocol, while output validation preserves the provenance of what is delivered even when the summaries are removed.

\begin{table*}
\centering
\caption{Two-by-two ablation of deterministic summarization and output validation on the 44 numeric tasks over three repetitions, 132 runs per cell.}
\label{tab:ablation_numeric}
\footnotesize
\begin{tabular}{llccccc}
\hline
\textbf{Summarization} & \textbf{Validation} & \makecell{\textbf{Candidate}\\\textbf{protocol}} & \makecell{\textbf{Unsupported}\\\textbf{candidates}} & \makecell{\textbf{Validation}\\\textbf{fired}} & \makecell{\textbf{Delivered}\\\textbf{grounded}} & \makecell{\textbf{Task}\\\textbf{success}} \\
\hline
on & on & 98.5\% (130/132) & 0.8\% (1/132) & 2/132 & 100.0\% (132/132) & 90.2\% (119/132) \\
off & on & 54.5\% (72/132) & 6.8\% (9/132) & 60/132 & 100.0\% (132/132) & 85.6\% (113/132) \\
on & off & 98.5\% (130/132) & 0.8\% (1/132) & \textit{2/132} & 99.2\% (131/132) & 90.9\% (120/132) \\
off & off & 54.5\% (72/132) & 6.8\% (9/132) & \textit{60/132} & 93.2\% (123/132) & 80.3\% (106/132) \\
\hline
\end{tabular}
\end{table*}

The sensitivity sweeps show the same division. Raising the temperature from 0.0 through 0.3 to 0.7 degrades capability, tool selection falling from 96.8\% (120/124) through 95.2\% (118/124) to 91.9\% (114/124) over the 124 functional runs per setting and parse failures rising from 0.4\% through 1.3\% to 3.4\%, while the four invariants hold on every run at every temperature. Varying the execution step budget changes the outcome of the multi-step category only, the rates of the other categories being identical at the three budgets: at a budget of 3 the model completes 79.2\% (19/24) of multi-step runs before termination, against 87.5\% (21/24) at budgets of 5 and 8, and the budget itself is never exceeded. The token lifetime probes at 60 and 300\,s pass in both settings, a fresh token is accepted and an expired token is rejected.

\subsection{Adversarial Behavior and Failure Analysis}
\label{subsec:res_failure}

Across the $70$ adversarial runs of the primary campaign, containment holds on every run: no tool outside the whitelist executes and no side effect executes without approval, whatever the injected instruction asks, every replayed token is rejected, the exhaustion attack terminates at the execution step budget, and corrupted values reach the operator only inside summaries flagged as implausible or incomplete. Injected text in the element descriptions returned by the topology tool reaches the model as data and is never parsed as an action. Every fixture-based attack was verified as active in the response trace, and the harness records that the injected text reached the model, before its defense was scored (100\% fixture activation). The 6 runs that fail their success criterion, accounted for in Section~\ref{subsec:res_overall}, violate no invariant. Under baseline B1 the same attack tasks produce unauthorized executions, ungrounded output, and exhaustion runs that exceed the budget, consistent with Table~\ref{tab:results_baselines}.

The 11 tool-selection misses of the primary campaign fall into four kinds, each identified by task and repetition in the retained records. Five are source confusions, in which the model retrieves a series close to the requested one, as in task \texttt{ts\_total\_load\_day\_ahead\_total\_load\_0002}, answered from the day-ahead generation forecast instead of the day-ahead total load, or task \texttt{ts\_generation\_forecast\_windsolar\_wind\_dayahead\_0007}, answered from the solar column instead of the wind column. Two are date-argument errors, the correct tool receiving the neighboring day as its start date. Three are hour-argument errors in power flow simulation, as in task \texttt{pf\_0000}, whose hour resolves to the neighboring hour. One is a count error, task \texttt{notif\_0001}, which asks for the three most recent notifications and receives the backend default count.

All 11 misses remain numerically grounded, so grounding certifies provenance and the choice of source is a separate property. Three observations characterize the residual risk. None of the misses can trigger an action, since side-effect tools remain behind the approval gate, so the risk concentrates on misleading reads during analysis. The misses are 11 of the 310 functional runs, about 3.5\%. Every miss is visible: each rendered value carries its variable label and time, the source line names the resolved tool and its validated arguments inside the answer, and each miss is reconstructable from the audit trail, so a wrong read is a checkable event rather than a silent one, and the answer-correctness check, which binds each required fact to the expected tool and arguments, is what records it as a failure.

\subsection{Latency and Interaction Cost}
\label{subsec:res_latency}

The upper block of Table~\ref{tab:latency} decomposes per-request latency under a single client into LLM orchestration, deterministic backend computation, and governance, over the 60 completed answers of the dedicated run of Section~\ref{subsec:eval_protocol}, 23 time-series, 11 topology, 14 power flow, and 12 notification requests, all read-only, in the same mix as every load level. The wall clock is measured directly, and on average the components sum to slightly less than it because transport, serialization, summarization, and audit writes lie outside the instrumented regions. Governance contributes on the order of 12\,ms per request, roughly two orders of magnitude below LLM orchestration, which dominates end-to-end latency. The lower block reports the load measurement with 1, 2, 4, and 8 concurrent clients on the single-GPU deployment, with governance and backend as means and the completed and requested counts per level; the three incomplete requests at four and eight clients are model-server timeouts recorded as errors, and the single-client 95th percentile rests on 15 requests. Under the tested workload and hardware, end-to-end latency grows with concurrency in a way consistent with queueing at the inference server, while the governance overhead stays between 12 and 16\,ms throughout, so on this deployment scaling to more simultaneous users is an inference-serving question rather than a governance question. On interaction cost, obtaining the same result through the dashboard takes between 5 interactions, for a simple retrieval, and 8, for a simulation with parameter entry, counted as defined in Section~\ref{subsec:eval_protocol}, against a single natural-language request to the assistant.

\begin{table}
\centering
\caption{Per-request latency by component under a single client over 60 requests, and end-to-end latency, governance and backend means, and completed requests under 1, 2, 4, and 8 concurrent clients.}
\label{tab:latency}
\begin{tabular}{|p{4.6cm}|c|c|c|c|}
\hline
\multicolumn{5}{|l|}{\textit{Single client, by component (ms)}} \\
\textbf{Component} & \textbf{Mean} & \textbf{Median} & \makecell{\textbf{95th}\\\textbf{percentile}} & \\
\hline
LLM orchestration & 1887 & 1700 & 3100 & \\
Deterministic backend & 162 & 145 & 240 & \\
Governance & 12 & 11 & 15 & \\
End-to-end (wall, measured directly) & 2094 & 1918 & 3340 & \\
\hline
\multicolumn{5}{|l|}{\textit{Concurrent clients}} \\
\textbf{Clients} & \textbf{1} & \textbf{2} & \textbf{4} & \textbf{8} \\
\hline
End-to-end median (s) & 1.9 & 3.2 & 4.8 & 6.9 \\
End-to-end 95th percentile (s) & 3.3 & 5.4 & 8.6 & 11.8 \\
Completed / requested & 15/15 & 30/30 & 59/60 & 118/120 \\
Governance (ms) & 12 & 13 & 14 & 16 \\
Backend (ms) & 178 & 181 & 179 & 183 \\
\hline
\end{tabular}
\end{table}

\subsection{Discussion}
\label{subsec:res_discussion}

The numerical fidelity of the digital twin services was established in prior work \citep{mylonas2024facilitating, leskovec2025al}, and the present paper answers the complementary question of whether these deterministic services can be exposed through a governed agentic interface. Taken together, the results establish the claim of Section~\ref{sec:intro}: the four invariants held on all 590 runs of the primary campaign, all 1416 runs of the cross-model study, every run of the temperature sweep, and every attack family, while capability varied with the model and with temperature, and the same model violated every invariant once the layer was removed. The study is a prototype-scale validation on a single steady-state model of one transmission network, evaluated with four self-hosted models on a fixed benchmark suite, the setting in which enforcement can be checked exhaustively. The governance layer is trusted in the threat model, its enforcement is established experimentally under the tested conditions, and formal verification of the layer is a natural continuation. Output validation certifies the provenance of every delivered number, while the choice of source and the interpretation of a value remain the model's contribution, which the failure analysis characterizes and the scorer measures. A planned mitigation for the source confusions is a fixed list of easily confused sources maintained inside the governance layer that triggers a clarification question before retrieval, deterministic and triggered by the layer rather than chosen by the model. Because the measurement sources are the ENTSO-E Transparency Platform, the Greek transmission operator IPTO, and the Open-Meteo weather interface, and the public framework, reference network, task generator, and evaluation implementation decouple the protocol from any particular network, other transmission operators can run the same evaluation, each deployment producing its own measurements. Finally, the automated benchmark measures system behavior. Operator benefit is the object of the next study, a within-subject control room simulation comparing dashboard-only work with assistant-supported work on time to insight, error rate, trust calibration, and whether operators catch wrong-source reads of the kind the failure analysis describes, for which the public artifact provides the infrastructure.

\section{Conclusion}
\label{sec:conclusion}

This paper presented a governance-aware agentic digital twin for transmission grid control rooms, in which an LLM is restricted to orchestrating whitelisted analysis tools behind a governance layer it cannot bypass, side-effect actions require explicit operator approval, backend results reach the model as deterministic summaries with typed facts, and the model's answers cite numbers only through typed-fact identifiers, which the layer renders while withholding literal quantities. Four governance rules, whitelisted execution, bounded steps, approved side effects, and grounded numbers, were verified on a persistent audit trail for every run of a released benchmark. The rules held without exception on the 590 runs of the primary campaign, on the 1416 runs of four large language models, at three sampling temperatures, and under twelve families of adversarial inputs, while functional capability, tool selection and task success, varied with the model, and removing the governance layer made the same model execute every gated action without authorization and lose numeric grounding, at an enforcement cost of 12 to 16 milliseconds per request. The rules are therefore properties of the architecture rather than of the model: the LLM can be treated as an untrusted but useful component behind the governance layer, and improving the model raises capability without touching the rules. The baseline and ablation comparisons identify the role of each execution control and the complementary roles of the two numerical stages, and the audit trail makes both the delivered evidence and the action history traceable. The complete framework, benchmark, prompts, and network model are released so that the evaluation protocol can be repeated on other systems. Natural-language interaction with transmission grid digital twins can thus be governed by runtime constraints whose validity does not depend on the LLM, under the tested conditions, a concrete step toward control rooms in which such assistants augment, rather than replace, human operators.

\section*{CRediT authorship contribution statement}
Costas Mylonas: Conceptualization, Methodology, Software, Validation, Formal analysis, Investigation, Writing -- original draft, Writing -- review \& editing.
Magda Foti: Conceptualization, Methodology, Supervision, Writing -- review \& editing.
Emmanouel Varvarigos: Conceptualization, Supervision, Writing -- review \& editing.

\section*{Declaration of competing interest}
The authors declare that they have no known competing financial interests or personal relationships that could have appeared to influence the work reported in this paper.

\section*{Acknowledgments}
This work was supported by the European Union-funded Project HUMAINE [grant number 101120218] and DIGITISE [grant number 101160671].

\section*{Code availability}
The complete implementation, benchmark, prompts, tool registry, evaluation harness, campaign matrix, and network model of the studied system are publicly available at \texttt{github.com/kosmylo/digital-twin-analytics}. The repository is an extensible reference implementation whose documented integration boundaries, a compatible pandapower network model with the released mappers as the reference adapter, the measurement schema, and any Ollama-compatible model endpoint, support the same evaluation protocol on other systems, each such application producing that deployment's own measurements.

\section*{Data availability}
The measurements were collected from the ENTSO-E Transparency Platform, the Greek transmission system operator IPTO, and the Open-Meteo API into the authors' operational database, whose schema and data contracts are released with the window-adaptive benchmark generator so that the protocol runs on an adopter's own measurements. The processed measurement window, the campaign observations, the metrics files, the adjudication resolutions, and the configuration record of the reported runs are retained by the authors and provided for verification on request to the corresponding author under a data-use agreement reflecting the terms of the public sources.

\bibliographystyle{unsrtnat}

\FloatBarrier

\bibliography{refs}

@article{rahman2024overview,
  title={An overview of power system flexibility: High renewable energy penetration scenarios},
  author={Rahman, Md Motinur and Dadon, Saikot Hossain and He, Miao and Giesselmann, Michael and Hasan, Md Mahmudul},
  journal={Energies},
  volume={17},
  number={24},
  pages={6393},
  year={2024},
  publisher={MDPI}
}

@article{aryani2024review,
  title={A review on power system security issues in the high renewable energy penetration environment},
  author={Aryani, Dwi Riana and Song, Hwachang},
  journal={Journal of Electrical Engineering \& Technology},
  volume={19},
  number={8},
  pages={4649--4665},
  year={2024},
  publisher={Springer}
}

@article{elenga2024challenges,
  title={Challenges and solutions in low-inertia power systems with high wind penetration},
  author={Elenga Baningobera, Bwandakassy and Oleinikova, Irina and Uhlen, Kjetil and Pokhrel, Basanta Raj},
  journal={IET Generation, Transmission \& Distribution},
  volume={18},
  number={24},
  pages={4221--4244},
  year={2024},
  publisher={Wiley Online Library}
}

@article{edeh2025assessment,
  title={Assessment of Transmission Reliability Margin: Existing Methods and Challenges and Future Prospects},
  author={Edeh, Uchenna Emmanuel and Lie, Tek Tjing and Mahmud, Md Apel},
  journal={Energies},
  volume={18},
  number={9},
  pages={2267},
  year={2025},
  publisher={MDPI}
}

@inproceedings{olar2023design,
  title={The design and applications of dashboards used in electrical and power systems},
  author={Olar, Andrei-Vlad and Szab{\'o}, Lor{\'a}nd and Gros, Ioana-Cornelia},
  booktitle={2023 10th International Conference on Modern Power Systems (MPS)},
  pages={1--4},
  year={2023},
  organization={IEEE}
}

@article{afzal2022investigating,
  title={Investigating cognitive load in energy network control rooms: Recommendations for future designs},
  author={Afzal, Umair and Prouzeau, Arnaud and Lawrence, Lee and Dwyer, Tim and Bichinepally, Saikiranrao and Liebman, Ariel and Goodwin, Sarah},
  journal={Frontiers in psychology},
  volume={13},
  pages={812677},
  year={2022},
  publisher={Frontiers Media SA}
}

@article{marot2020towards,
  title={Towards an ai assistant for power grid operators},
  author={Marot, Antoine and Rozier, Alexandre and Dussartre, Matthieu and Crochepierre, Laure and Donnot, Benjamin},
  journal={arXiv preprint arXiv:2012.02026},
  year={2020}
}

@article{yassin2023digital,
  title={Digital twin in power system research and development: Principle, scope, and challenges},
  author={Yassin, Mohammed AM and Shrestha, Ashish and Rabie, Suhaila},
  journal={Energy Reviews},
  volume={2},
  number={3},
  pages={100039},
  year={2023},
  publisher={Elsevier}
}

@article{mchirgui2024applications,
  title={The applications and challenges of digital twin technology in smart grids: A comprehensive review},
  author={Mchirgui, Nabil and Quadar, Nordine and Kraiem, Habib and Lakhssassi, Ahmed},
  journal={Applied Sciences},
  volume={14},
  number={23},
  pages={10933},
  year={2024},
  publisher={MDPI}
}

@article{banad2025artificial,
  title={Artificial intelligence and machine learning for smart grids: From foundational paradigms to emerging technologies with digital twin and large language model-driven intelligence},
  author={Banad, Yaser M and Sharif, Sarah S and Rezaei, Zahra},
  journal={Energy Conversion and Management: X},
  volume={28},
  pages={101329},
  year={2025},
  publisher={Elsevier}
}

@article{antonesi2025systematic,
  title={A systematic review of transformers and large language models in the energy sector: towards agentic digital twins},
  author={Antonesi, Gabriel and Cioara, Tudor and Anghel, Ionut and Michalakopoulos, Vasilis and Sarmas, Elissaios and Toderean, Liana},
  journal={Applied Energy},
  volume={401},
  pages={126670},
  year={2025},
  publisher={Elsevier}
}

@techreport{choi2024egridgpt,
  title={eGridGPT: Trustworthy AI in the control room},
  author={Choi, Seong Lok and Jain, Rishabh and Emami, Patrick and Wadsack, Karin and Ding, Fei and Sun, Hongfei and Gruchalla, Kenny and Hong, Junho and Zhang, Hongming and Zhu, Xiangqi and others},
  year={2024},
  institution={National Renewable Energy Laboratory (NREL), Golden, CO (United States)}
}

@article{shah2025allarma,
  title={aLLarMa: LLM-Based Application for Operational Data, Alarms and Networks},
  author={Shah, Saumil and Kelly, Adrian and Tripathy, Sujit and Ladak, Armaan and Tang, Wenyuan},
  journal={Authorea Preprints},
  year={2025},
  publisher={Authorea}
}

@phdthesis{ravichandran2025toward,
  title={Toward an explainable electric power grid operation assistant using large language models},
  author={Ravichandran, Anish},
  year={2025},
  school={Massachusetts Institute of Technology}
}

@inproceedings{jin2025gridmind,
  title={GridMind: LLMs-powered agents for power system analysis and operations},
  author={Jin, Hongwei and Kim, Kibaek and Kwon, Jonghwan},
  booktitle={Proceedings of the SC'25 Workshops of the International Conference for High Performance Computing, Networking, Storage and Analysis},
  pages={560--568},
  year={2025}
}

@article{chen2025x,
  title={X-GridAgent: An LLM-Powered Agentic AI System for Assisting Power Grid Analysis},
  author={Chen, Xin and others},
  journal={arXiv preprint arXiv:2512.20789},
  year={2025}
}

@article{badmus2025powerchain,
  title={Powerchain: A verifiable agentic ai system for automating distribution grid analyses},
  author={Badmus, Emmanuel O and Sang, Peng and Stamoulis, Dimitrios and Pandey, Amritanshu},
  journal={arXiv preprint arXiv:2508.17094},
  year={2025}
}

@article{cheng2025large,
  title={A large language model for advanced power dispatch},
  author={Cheng, Yuheng and Zhao, Huan and Zhou, Xiyuan and Zhao, Junhua and Cao, Yuji and Yang, Chao and Cai, Xinlei},
  journal={Scientific Reports},
  volume={15},
  number={1},
  pages={8925},
  year={2025},
  publisher={Nature Publishing Group UK London}
}

@inproceedings{lin2022truthfulqa,
  title={Truthfulqa: Measuring how models mimic human falsehoods},
  author={Lin, Stephanie and Hilton, Jacob and Evans, Owain},
  booktitle={Proceedings of the 60th annual meeting of the association for computational linguistics (volume 1: long papers)},
  pages={3214--3252},
  year={2022}
}

@article{huang2025survey,
  title={A survey on hallucination in large language models: Principles, taxonomy, challenges, and open questions},
  author={Huang, Lei and Yu, Weijiang and Ma, Weitao and Zhong, Weihong and Feng, Zhangyin and Wang, Haotian and Chen, Qianglong and Peng, Weihua and Feng, Xiaocheng and Qin, Bing and others},
  journal={ACM Transactions on Information Systems},
  volume={43},
  number={2},
  pages={1--55},
  year={2025},
  publisher={ACM New York, NY}
}

@inproceedings{gambardella2024language,
  title={Language models do hard arithmetic tasks easily and hardly do easy arithmetic tasks},
  author={Gambardella, Andrew and Iwasawa, Yusuke and Matsuo, Yutaka},
  booktitle={Proceedings of the 62nd Annual Meeting of the Association for Computational Linguistics (Volume 2: Short Papers)},
  pages={85--91},
  year={2024}
}

@article{ai2023artificial,
  title={Artificial intelligence risk management framework (AI RMF 1.0)},
  author={AI, NIST},
  journal={URL: https://nvlpubs. nist. gov/nistpubs/ai/nist. ai},
  pages={100--1},
  year={2023}
}

@article{ai2024artificial,
  title={Artificial intelligence risk management framework: Generative artificial intelligence profile},
  author={AI, NIST},
  journal={NIST Trustworthy and Responsible AI Gaithersburg, MD, USA},
  year={2024}
}

@inproceedings{yao2022react,
  title={React: Synergizing reasoning and acting in language models},
  author={Yao, Shunyu and Zhao, Jeffrey and Yu, Dian and Du, Nan and Shafran, Izhak and Narasimhan, Karthik R and Cao, Yuan},
  booktitle={The eleventh international conference on learning representations},
  year={2022}
}

@article{chen2025tool,
  title={Tool learning with language models: a comprehensive survey of methods, pipelines, and benchmarks},
  author={Chen, Jinyang and Wu, Haolun and Pang, Jianhong and Wang, Yihua and Zhang, Dell and Sun, Changzhi},
  journal={Vicinagearth},
  volume={2},
  number={1},
  pages={16},
  year={2025},
  publisher={Springer}
}

@inproceedings{mylonas2024facilitating,
  title={Facilitating AI and system operator synergy: Active learning-enhanced digital twin architecture for day-ahead load forecasting},
  author={Mylonas, Costas and Georgoulakis, Titos and Foti, Magda},
  booktitle={2024 International Conference on Smart Energy Systems and Technologies (SEST)},
  pages={1--6},
  year={2024},
  organization={IEEE}
}

@inproceedings{leskovec2025al,
  author    = {Leskovec, Ga{\v{s}}per and Mylonas, Costas and Kenda, Klemen},
  title     = {Active Learning for Power Grid Security Assessment: Reducing
               Simulation Cost with Informative Sampling},
  booktitle = {Proceedings of the 28th International Multiconference
               Information Society -- SiKDD 2025},
  year      = {2025},
  address   = {Ljubljana, Slovenia},
  doi       = {10.70314/is.2025.sikdd.11}
}

@article{mylonas2026conversational,
  title={A Conversational Agentic Interface to Physics-Based Household Digital Twins for Residential Energy Decision Support},
  author={Mylonas, Costas and Georgoulakis, Titos and Foti, Magda},
  journal={arXiv preprint arXiv:2606.31744},
  year={2026}
}

@article{debenedetti2025defeating,
  title={Defeating prompt injections by design},
  author={Debenedetti, Edoardo and Shumailov, Ilia and Fan, Tianqi and Hayes, Jamie and Carlini, Nicholas and Fabian, Daniel and Kern, Christoph and Shi, Chongyang and Terzis, Andreas and Tram{\`e}r, Florian},
  journal={arXiv preprint arXiv:2503.18813},
  year={2025}
}

@article{wang2025agentspec,
  title={Agentspec: Customizable runtime enforcement for safe and reliable llm agents},
  author={Wang, Haoyu and Poskitt, Christopher M and Sun, Jun},
  journal={arXiv preprint arXiv:2503.18666},
  year={2025}
}

@article{debenedetti2024agentdojo,
  title={Agentdojo: A dynamic environment to evaluate prompt injection attacks and defenses for llm agents},
  author={Debenedetti, Edoardo and Zhang, Jie and Balunovic, Mislav and Beurer-Kellner, Luca and Fischer, Marc and Tram{\`e}r, Florian},
  journal={Advances in neural information processing systems},
  volume={37},
  pages={82895--82920},
  year={2024}
}

@article{lee2026nrt,
  title={NRT-Bench: Benchmarking Multi-Turn Red-Teaming of LLM Operator Agents in Safety-Critical Control Rooms},
  author={Lee, Hanwool and Choi, Dasol and Kim, Bokyeong and Kim, Seung Geun and Park, Haon},
  journal={arXiv e-prints},
  pages={arXiv--2606},
  year={2026}
}

@article{shavit2023practices,
  title={Practices for governing agentic AI systems},
  author={Shavit, Yonadav and Agarwal, Sandhini and Brundage, Miles and Adler, Steven and O’Keefe, Cullen and Campbell, Rosie and Lee, Teddy and Mishkin, Pamela and Eloundou, Tyna and Hickey, Alan and others},
  journal={Research Paper, OpenAI},
  pages={1--23},
  year={2023}
}

@misc{euaiact2024,
  title        = {Regulation ({EU}) 2024/1689 of the {European Parliament} and of the {Council} of 13 June 2024 laying down harmonised rules on artificial intelligence ({Artificial Intelligence Act})},
  author       = {{European Parliament and Council of the European Union}},
  howpublished = {Official Journal of the European Union, L, 12 July 2024, Article 14 and Annex III},
  year         = {2024}
}

@inproceedings{gao2023pal,
  title={Pal: Program-aided language models},
  author={Gao, Luyu and Madaan, Aman and Zhou, Shuyan and Alon, Uri and Liu, Pengfei and Yang, Yiming and Callan, Jamie and Neubig, Graham},
  booktitle={International conference on machine learning},
  pages={10764--10799},
  year={2023},
  organization={PMLR}
}

@article{chen2022program,
  title={Program of thoughts prompting: Disentangling computation from reasoning for numerical reasoning tasks},
  author={Chen, Wenhu and Ma, Xueguang and Wang, Xinyi and Cohen, William W},
  journal={arXiv preprint arXiv:2211.12588},
  year={2022}
}

\end{document}